\documentclass[aps,showpacs,twocolumn,floats,superscriptaddress]{revtex4}
\usepackage{hyperref}
\usepackage{graphicx}
\usepackage{color}

\usepackage{amsmath}

\def\NOT(#1,#2){\OneQubitGate(#1,#2){$X$}}

\begin{document}

\title{Experimental Quantum Simulation of Entanglement in Many-body Systems}
\author{
Jingfu Zhang,$^1$ Tzu-Chieh Wei$^{1,3}$, and Raymond Laflamme$^{1,2}$\\
\it {$^1$Institute for Quantum Computing and Department of
Physics,
University of Waterloo, Waterloo, Ontario, Canada N2L 3G1\\
$^2$Perimeter Institute for Theoretical Physics, Waterloo, Ontario, N2J 2W9,
Canada\\
$^3$ Department of Physics and Astronomy, University of British Columbia,
Vancouver, BC V6T 1Z1, Canada}}
\date{\today}

\begin{abstract}
 We employ a nuclear magnetic resonance (NMR)
quantum information processor to simulate the ground state of an XXZ
spin chain and measure its NMR analog of entanglement, or
pseudo-entanglement. The observed pseudo-entanglement for a
small-size system already displays singularity, a signature which is
qualitatively similar to that in the thermodynamical limit across
quantum phase transitions, including an infinite-order critical
point. The experimental results illustrate a successful approach to
investigate quantum correlations in many-body systems using quantum
simulators.
\end{abstract}
\pacs{03.67.Ac, 03.67.Lx, 75.10.Pq} \maketitle

  Entanglement,  delineated as the non-local  correlation, is one ``spooky''
  characteristic trait of quantum
mechanics \cite{EPR}. The famous dispute between Bohr and Einstein
on the fundamental question of quantum mechanics,
Schr\"{o}dinger-cat paradox,
 and the
transitions from quantum to classical worlds essentially involve
entanglement. Recent development of quantum information rekindles
the interest in entanglement, more as a
 possible resource for information
processing ~\cite{NielsenChuang}. 
Various methods have been proposed to characterize entanglement
qualitatively and quantitatively~\cite{Entanglement2}. One
immediate application of the entanglement is the investigation of
quantum phase transitions
(QPTs)~\cite{reviewQPT,Concurrence,Sachdev08} in many-body
systems, which occur at zero temperature (T=0 K), where the
transitions are driven by quantum fluctuations and the
ground-state wave function is expected to develop drastic change.
The entanglement properties extend and complement the traditional
statistical-physical methods for QPTs, such as the correlation
functions and low-lying excitation spectra. However, how to
describe and measure entanglement in many-body systems is still a
challenging task in both theoretical and experimental
aspects~\cite{many,detect}. Most schemes for directly measuring
entanglement focus on the entanglement between two
qubits~\cite{Horodecki03,WSDMB06}.  Although the degree of
entanglement for a medium-size or lager system can in principle be
probed~\cite{GRW07,detect}, it has not been experimentally
measured directly.

In contrast to classical approaches, quantum simulators
~\cite{Feynman} provide a promising approach for investigating
many-body systems and  enable one to efficiently simulate other
quantum systems by actively controlling and manipulating a certain
quantum system, and to test, probe and unveil new physical
phenomena. One interesting aspect is to simulate the ground states
of many-body systems, where usually rich phases can exist, such as
ferromagnetism, superfluidity,
 and quantum Hall effect,
just to name a few.
 In this Letter we experimentally simulate the ground state of an XXZ spin chain~\cite{XXZbook} in
a liquid-state nuclear magnetic resonance (NMR) quantum information
processor~\cite{rmp} and directly  measure a global multipartite
entanglement ---the geometric entanglement (GE)
~\cite{WeiPRA03,WeiPRA05} in a version of NMR analog, or
pseudo-entanglement. Exploiting the probed behavior of GE, we
identify two QPTs, which in the thermodynamic limit correspond to
the first- and $\infty$-orders, respectively. In the $\infty$-order
QPT, also known as Kosterlitz-Thouless (KT)
transition~\cite{KT,PRA81032334}, the ground-state energy is {\it
not} singular. Consequently the detection of the critical point in
the KT transition may pose a challenge for the correlation-based
approaches~\cite{Concurrence,Wu}, which rely on the singularity of
ground-state energy.
 Surprisingly, the GE turns out to be non-analytical but of
different types of singularity at the first- and $\infty$-order
transitions~\cite{OrusWei}.
 Remarkably, the qualitative features
of the ground-state entanglement in the thermodynamic limit
displayed near both transitions persist even for a small-size
system, on which our experiment is performed.

The GE of a pure many-spin quantum state $|\Phi\rangle$ is
captured by the maximal overlap \cite{WeiPRA05,WeiPRA03}
    $\Lambda_{\max}\equiv \max_{\Psi} |\langle \Psi|\Phi\rangle|$,
and is defined as
   $ E_{\log_{2}}=-\log_{2}\Lambda^{2}_{\max}$,
 where
$|\Psi\rangle\equiv\bigotimes_{i=1}^{N}|\psi^{(i)}\rangle$ denotes
all product (i.e. unentangled) states of the $N$-spin system.
From the point of view of local measurements, the GE  is
essentially (modulo a logarithmic function) the maximal
probability that can be achieved by a local projective measurement
on every site, and the closest product state signifies the optimal
measurement setting. The GE has been employed 
 to study QPTs~\cite{WeiPRA05,GEQPT1}, local state
discrimination~\cite{StateDis} and 
entanglement as computational resources~\cite{TooEntangled}.

The XXZ spin chain 
is described by the Hamiltonian
\begin{equation}\label{hamXXZ}
    H_{XXZ}=\sum_{i=1}^{N} (X_{i}X_{i+1}+Y_{i}Y_{i+1}+\gamma
    Z_{i}Z_{i+1})
\end{equation}
where $X_{i}$, $Y_{i}$, $Z_{i}$ denote the Pauli matrices with $i$
indicating the spin location, and $\gamma$ is the control parameter
for QPTs. We use the periodic boundary condition with $N+1 \equiv
1$. The XXZ chain can be exactly solved by the so-called Bethe
Ansatz and exhibits rich phase diagrams in the ground
state~\cite{XXZbook}. In the thermodynamic limit, for $\gamma <-1$,
 the system has the
ground state with the ferromagnetic (FM) Ising phase. At
$\gamma=-1$ a first-order QPT occurs. For $-1<\gamma \leq 1$, the
system is in a gapless phase or XY-like phase. At $\gamma=1$ there
is an $\infty$-order or a KT transition~\cite{KT}, where the
ground-state energy, however, is analytic across the transition,
and so is any correlation function. For $\gamma
> 1$, the system is in the N\'eel-like antiferromagnetic (AFM) phase. The ground state is asymptotically doubly degenerate.
However, the excitations above the ground space have a gap. For
$\gamma \gg 1$, the ground state takes a N\'eel or Ising AFM state
i.e.,
$|...10101010...\rangle$, where 
$|0\rangle\equiv |\uparrow\rangle$ and
$|1\rangle\equiv|\downarrow\rangle$.

In the XXZ chain (\ref{hamXXZ}), the  GE  displays a jump across
$\gamma=-1$ but a cusp (i.e., the derivative is discontinuous)
across $\gamma=1 $~\cite{OrusWei}. 
Both features are present for small-size systems, as well as in the
thermodynamic limit. For $\gamma< -1$ the GE is essentially zero.
Regardless of the system size (as long as it is even), for $-1
<\gamma \le 1$, the closest product state is found to be
$|+-+-...\rangle$,
whereas for $\gamma \ge 1$, the closest product state is found to be
$|0101...\rangle$, where $|\pm\rangle
\equiv(|0\rangle\pm|1\rangle)/\sqrt{2}$ \cite{note1}.
 Right at the KT point $\gamma=1$, because of
rotational symmetry, the closest product states are
$|\phi\,\phi^\perp\phi\,\phi^\perp...\rangle$, where
$|\phi\rangle$ and $|\phi^\perp\rangle$ are any arbitrary
orthonormal qubit states. The singular behavior of ground-state GE
can be used to probe the KT transition, and there is no need to
know the low-lying spectrum.


\begin{figure}
\includegraphics[width=3.5in]{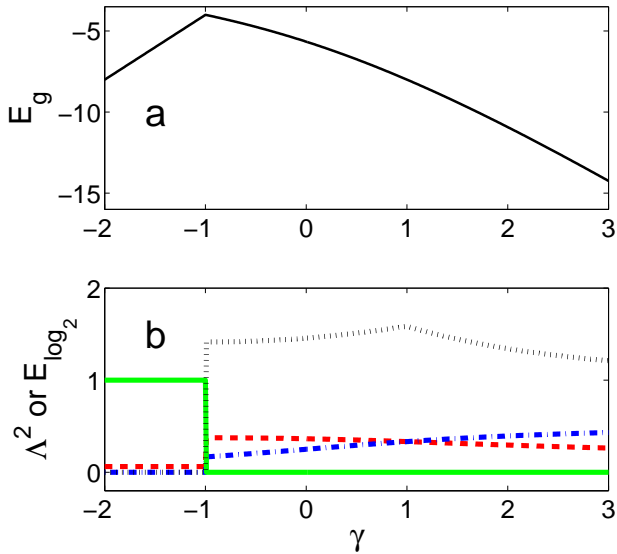}
\caption{(Color online). Theoretical results in the 4-spin XXZ
 chain. (a) Energy level of the ground state. (b) The overlap
square $\Lambda^{2}_{i}(\gamma)$ and entanglement
$E_{\log_{2}}(\gamma)$. The  solid, dashed and  dash-dotted curves
show 
$\Lambda^{2}_{i}(\gamma)$ for $|\Psi_{1}\rangle=|1111\rangle$,
$|\Psi_{2}\rangle=|+-+-\rangle$ and
$|\Psi_{3}\rangle=|0101\rangle$, respectively.
$E_{\log_{2}}(\gamma)$ is shown as the dotted curve. The jump at
$\gamma=-1$ and the cusp at $\gamma=1$ in
 $E_{\log_{2}}(\gamma)$ indicate the
transition points for the QPTs, with the first- and $\infty$-
orders, respectively.} \label{figth}
\end{figure}

In implementation we use a 4-spin chain. 
The entanglement features pertinent to the QPTs in the
thermodynamic limit will survive. 
The ground-state energy and wave function of the 4-spin chain are
represented as (see Supplemental Material)
\begin{equation}\label{wholegroundE}
    E_{g}=\begin{cases}
    4\gamma  & (\gamma<-1)\\
-2\gamma-2\sqrt{\gamma^{2}+8} & (\gamma
>-1)
   \end{cases},
\end{equation}
\begin{equation}\label{wholeground}
    |g\rangle=\begin{cases}
    |1111\rangle  & (\gamma<-1)\\
|\phi_{1}\rangle \cos\alpha + |\phi_{2}\rangle\sin\alpha & (\gamma
>-1)
   \end{cases},
\end{equation}
where $\alpha\in(-\pi/2,0)$ is given via
 $ \tan(2\alpha)=-2\sqrt{2}/\gamma$,
 and
  $|\phi_{1}\rangle\equiv(|0101\rangle+|1010\rangle)/\sqrt{2}$,
    $|\phi_{2}\rangle
    \equiv(|1100\rangle+|0011\rangle+|1001\rangle+|0110\rangle)/2$.
Fig. \ref{figth} (a) shows $E_{g}$ as a function of $\gamma$. One
should  note that $E_{g}$ is continuous at $\gamma=-1$ while the
ground state $|g\rangle$ is discontinuous.

  In order to obtain the ground-state GE, one needs to search for the closest product state $|\Psi^*\rangle$ ~\cite{OrusWei}.
 In fact, we can choose the product states
  \begin{equation}\label{closestproduct}
    |\Psi^*(\gamma)\rangle=\begin{cases}
   \,\,|\Psi_{1}\rangle \equiv|1111\rangle & (\gamma<-1)\\
\,\,|\Psi_{2}\rangle \equiv|+-+-\rangle & (-1<\gamma <1)\\
\,\,|\Psi_{3}\rangle \equiv |0101\rangle & (\gamma >1)
   \end{cases}
\end{equation}
 to obtain the corresponding entanglement in the respective range of
 $\gamma$.
To anticipate the experimental procedure, we shall measure the
ground-state overlap listed as 
\begin{equation}\label{overlap1}
    \Lambda_{1}(\gamma)=\langle\Psi_{1}|g\rangle=\begin{cases}
    1 & (\gamma<-1)\\
0 & (\gamma >-1)
   \end{cases}
\end{equation}
\begin{equation}\label{overlap2}
    \Lambda_{2}(\gamma)=\langle\Psi_{2}|g\rangle=\begin{cases}
    \frac{1}{4} & (\gamma<-1)\\
\frac{\sqrt{2}}{4}\cos\alpha-\frac{1}{2}\sin\alpha & (\gamma >-1)
   \end{cases}
\end{equation}
\begin{equation}\label{overlap3}
    \Lambda_{3}(\gamma)=\langle\Psi_{3}|g\rangle=\begin{cases}
    0 & (\gamma<-1)\\
\frac{1}{\sqrt{2}}\cos\alpha & (\gamma >-1)
   \end{cases}.
\end{equation}
From Eqs. (\ref{overlap2}) and (\ref{overlap3}), one finds that
$\Lambda_{2}(\gamma)$ and  $\Lambda_{3}(\gamma)$ 
cross at $\gamma=1$. Fig. \ref{figth} (b) shows the theoretical
prediction for  $\Lambda^{2}_{i}(\gamma)$ ($i=1$, $2$, $3$) and
the entanglement $E_{\log_{2}}$.
 The jump in the entanglement at $\gamma=-1$ and the cusp at $\gamma=1$
  signify the two QPT points.


\begin{figure} 
\includegraphics[width=3.5in]{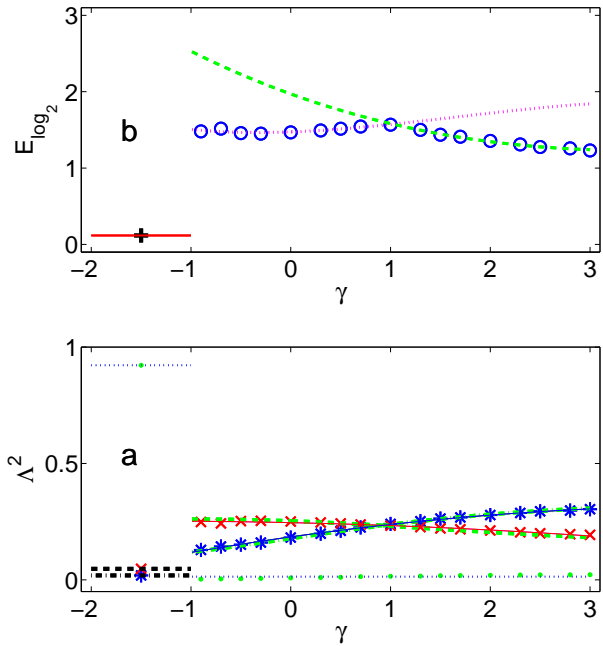} 
\caption{(Color online).  Experimentally measured
$\Lambda^{2}_{i}(\gamma)$ (a) and  $E_{\log_{2}}$ (b) for various
$\gamma$. In figure (a), the experimental data are shown as
''$\cdot$'', ''$\times$'' and ''*'' for $\Lambda^{2}_{i} =
|\langle \Psi_{i}|g\rangle|^{2}$, corresponding to
$|\Psi_{1}\rangle$, $|\Psi_{2}\rangle$ and $|\Psi_{3}\rangle$. The
measured $\Lambda^{2}_{i}(\gamma)$ for
$\gamma<-1$ are 
indicated as the dotted, dashed
and dash-dotted lines. 
In the region $\gamma>-1$, $\Lambda^{2}_{1}(\gamma)$
can be fitted  as 
 the dotted line.
 Through fitting the points for
$\Lambda^{2}_{2}(\gamma)$ and $\Lambda^{2}_{3}(\gamma)$ using
polynomial functions, we obtain the
 two solid curves that cross at point  $\gamma=0.92$, close to the
theoretical point at $\gamma=1$. The thick dashed and dash-dotted
curves show the fitting results using the theoretical
$\Lambda^{2}_{2}(\gamma)$ and $\Lambda^{2}_{3}(\gamma)$ by
introducing decay factors $0.69$ and $0.71$, respectively.
In figure (b), in range $\gamma<-1$, $E_{\log_{2}}$ is shown as
''+''. For $\gamma>-1$,
  we rescale
the measured $\Lambda^{2}_{2}(\gamma)$ and
$\Lambda^{2}_{3}(\gamma)$ as $\Lambda^{2}_{2}(\gamma)/0.69$ and
$\Lambda^{2}_{3}(\gamma)/0.71$, respectively.
The dotted and dashed curves that cross at $\gamma=1.02$ show the
fitting results of the rescaled data using polynomial functions.
The expected $E_{\log_{2}}$ after rescaling is indicated by
''$\circ$''. } \label{figres}
\end{figure}

\begin{figure}
\includegraphics[width=3.5in]{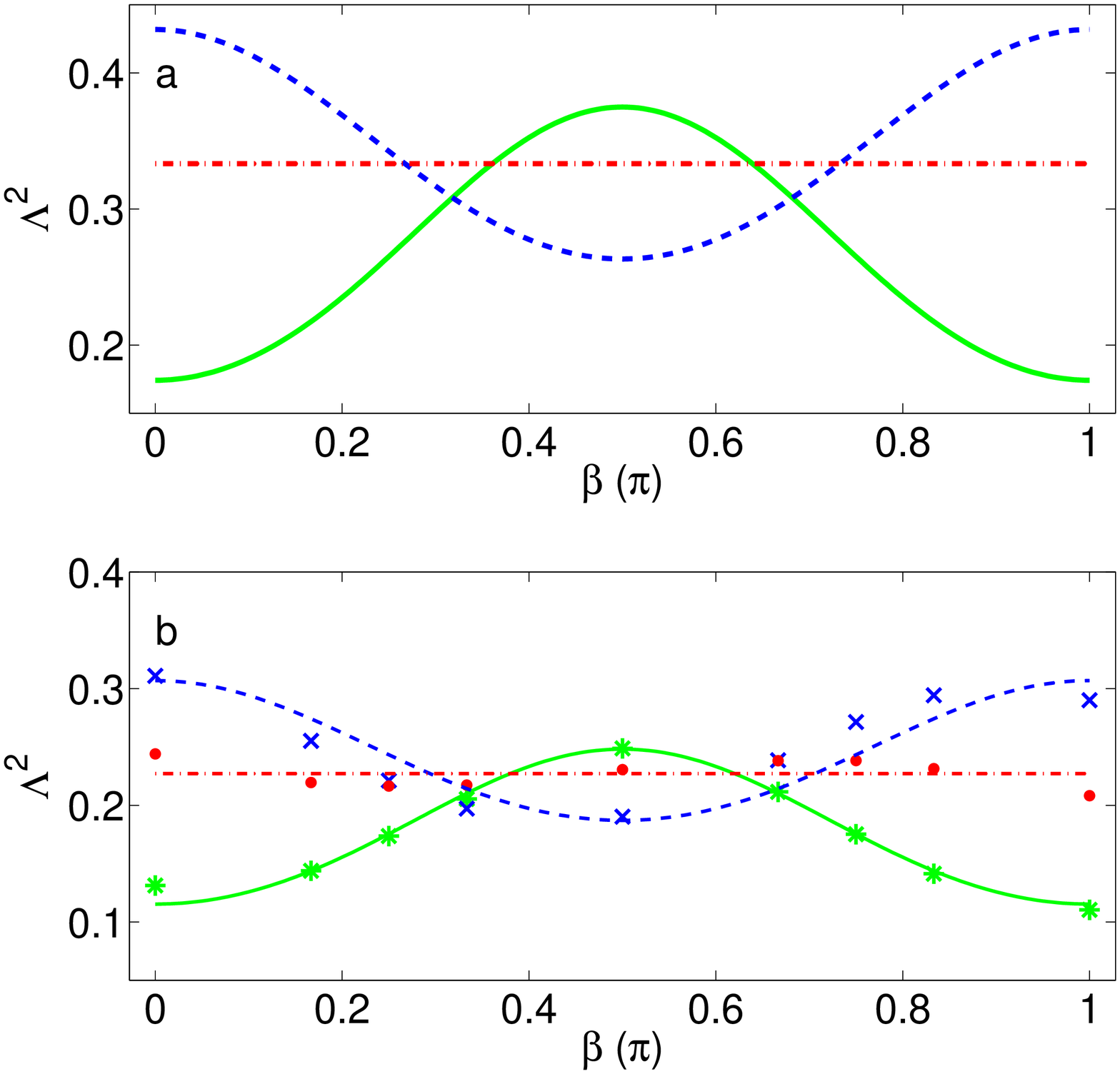} 
\caption{(Color online).  Theoretical (a) and experimentally
measured (b) $\Lambda^{2}$ for various product states created as
$U_{p}(\beta)|0101\rangle$ [see Eq. (\ref{special})] . Three
ground states for $\gamma=-0.9$, $1$ and $3$ are chosen and the
corresponding $\Lambda^{2}$ are shown as the the solid,
dash-dotted and  dashed curves in figure (a), respectively.
The experimental data are shown as ''*'', ''$\cdot$'' and
''$\times$'' for $\gamma=-0.9$, $1$ and $3$ in figure (b),
respectively. In comparison with the theoretical values, they can
be fitted as $0.66\Lambda^{2}$,
 $0.68\Lambda^{2}$ and  $0.71\Lambda^{2}$, shown as the solid,
 dash-dotted and  dashed curves.} \label{figres26}
\end{figure}

  In experiment, we choose the four carbons in crotonic acid
  \cite{crot}
  dissolved in d6-acetone as the four qubits.
  We generate the ground states using quantum
networks 
and implement the quantum gates by GRAPE pulses~\cite{GRAPE}.
Various ground-states can be created by varying the single spin
rotations that can be easily implemented (see Supplemental
Material). In principle, one can employ an iterative method to
experimentally measure the GE of the ground state (see
Supplemental Material). To demonstrate the proof-of-principle
simulation of quantum entanglement, instead, we first measure the
overlap of the ground state with several product
states~(\ref{closestproduct}), which contain the closest product
states. From the measurement with the already known closest
product states, we can obtain the ground-state GE. Next, to show
that the obtained results are the optimum, we vary the product
states to test the optimality.

   The experimentally measured $\Lambda_{i}^{2}(\gamma)$
for various $\gamma$ are shown in Fig. \ref{figres} (a).
The measured $\Lambda^{2}_{i}(\gamma)$ for
$\gamma<-1$ are $0.92$, $0.048$ and $0.019$, indicated as the
dotted, dashed and dash-dotted lines. The corresponding
theoretical values are $1$, $1/16$ and $0$, respectively. In the
region $\gamma>-1$, $\Lambda^{2}_{1}(\gamma)$ can be fitted  as
$\Lambda^{2}_{1}(\gamma) = 0.014$, shown as the dotted line,
corresponding to $0$ in theory.

We perform  polynomial fits to the measured
$\Lambda^{2}_{2}(\gamma)$ and $\Lambda^{2}_{3}(\gamma)$, and
obtain the
 two solid curves  that cross at  the point  $\gamma=0.92$,
 which is very close to the
theoretically predicted transition point at $\gamma=1$. The
discrepancy between experiment and theory mainly comes from the
different experimental errors in measuring $\Lambda^{2}_{2}(\gamma)$
and $\Lambda^{2}_{3}(\gamma)$. The jump at $\gamma=-1$ and the cusp
at $\gamma=0.92$ reflect the different types of
 QPT points.

  In order to faithfully estimate the performance of the experiment
in measuring $\Lambda^{2}_{2}(\gamma)$ and
$\Lambda^{2}_{3}(\gamma)$ in the range $\gamma>-1$, we introduce
two decay factors $\alpha_2$ and $\alpha_3$ to fit the
experimental data as $[\Lambda^{2}_{2,3}(\gamma)]_{{\rm
exp}}=\alpha_{2,3}[\Lambda^{2}_{2,3}(\gamma)]_{{\rm theory}}$,
shown as the thick dashed and dash-dotted curves in Fig.
\ref{figres} (a) with the best scale-factors as $\alpha_2 = 0.69$
and $\alpha_3 =0.71$, respectively. The difference between the
decay factors comes from the different operations in measuring
$\Lambda^2_2(\gamma)$ and $\Lambda^2_3(\gamma)$. In Fig.
\ref{figres} (b), we exploit the decay factors to rescale
experimental values of $[\Lambda^{2}_{2,3}]_{\rm
exp}/\alpha_{2,3}$, from which we obtain the expected values of
pseudo-entanglement shown as ''$\circ$''. The rescaled $-\log_2
([\Lambda^{2}_{2,3}]_{\rm exp}/\alpha_{2,3})$ can be fitted as the
dotted and dashed curves that cross at $\gamma=1.02$.

In principle, we do not need to know the closest product states in
order to measure the entanglement. In the Supplemental Material,
we describe an iterative procedure to search for them and this
procedure can be implemented in experiment. For proof-of-principle
demonstration of the optimality experimentally, we
  simplify the procedure and vary the product states
$|\Psi(\beta)\rangle$ by
\begin{equation}\label{special}
   |\Psi(\beta)\rangle=U_{p}(\beta)|0101\rangle,
\end{equation}
where
$U_{p}(\beta)=\bigotimes_{j=1}^{4} e^{-i\beta Y_{j}/2}$,
and experimentally measure $\Lambda^{2} =
|\langle\Psi(\beta)|g\rangle|{^2}$ for various $\beta$ at three
different locations of the phase diagram, corresponding to
$\gamma=-0.9$, $1$, and $3$, respectively.
The theoretical and experimental results are shown in
Figs.~\ref{figres26}(a) and (b), respectively. The experimental data
 are compared to the theoretical values of
$0.66\Lambda^{2}$,
 $0.68\Lambda^{2}$ and  $0.71\Lambda^{2}$, shown in Fig.~\ref{figres26} (b). One  finds that the
maximum of $\Lambda^{2}$ occurs at $\beta=\pi/2$ and $0$
 for $\gamma=-0.9$ and
$3$, respectively. These correspond to the  respective closest
product states, $|\Psi_{2}\rangle=|+-+-\rangle$ and
$|\Psi_{3}\rangle = |0101\rangle$, predicted theoretically.
Remarkably, for $\gamma=1$, where the $\infty$-order QPT occurs,
$\Lambda^{2}$ is a constant independent of  $\beta$, as we have
expected and noted
earlier. 
This also means that arbitrary states prepared by
Eq.~(\ref{special}) can be chosen to measure the entanglement at
$\gamma=1$, and this gives additional confirmation that the created
state at the KT point is rotationally invariant.

 The experiment duration of the preparation of the
ground states for $\gamma >-1$ is about 160 ms,
which is non-negligible (about 17\%) 
compared to the coherence time $T_{2}$. 
Consequently the decay of the signals due to the limitation of
coherence time is one of main sources of errors. Additionally
the imperfection of pulses 
and
inhomogeneities of magnetic fields 
 also contribute to errors.
The deviations of the experimental data from the theoretical
fitting in Fig.~\ref{figres26} (b) represent the effects of the
errors that depend on the rotation angles, or the product states.
In particular the fluctuation of the data for $\gamma=1$ in
Fig.~\ref{figres26} (b) confirms the explanation for the shift of
the measured cusp in Fig. \ref{figres} (a).


  In conclusion we demonstrate the non-analytic properties of many-body systems
in a quantum simulator using NMR. The QPTs with first- and
$\infty$-orders in the XXZ spin chain are detected by directly
measuring the pseudo-entanglement of the ground states created by
quantum gates. 
An alternative approach for creating ground states would be via
adiabatic evolution~\cite{Peng09}.  Our preliminary numerical
analysis indicates that ground states for $\gamma
>-1$ can be approximately  generated with high fidelity (e.g. $>0.998$) by the adiabatic evolution
from the ground state at a large $\gamma$. The experimental
implementation is a possible future direction.



We thank O. Moussa and R. Or\'us for helpful discussions. This
work was supported by CIFAR (R.L.), NSERC (J.-F.Z., R.L. and
T.-C.W.), MITACS (T.-C.W.), SHARCNET (R.L.), and QuantumWorks
(R.L.).

\appendix

\section*{{\large Supplemental Material}}

\section{Method for computing and measuring the maximal overlap
by iteration} \label{sec:iteration}
Here we describe an iterative
method to compute the maximal overlap, of which the motivation
comes from the density-matrix-renormalization-group (DMRG) or
matrix-product-state (MPS) variational method~\cite{mps}. This
method can not only be implemented numerically, but can also be
carried out experimentally. To compute the maximal overlap for the
state $|g\rangle$ with respect to product states
$|\Psi\rangle\equiv\bigotimes_{i=1}^{N}|\psi^{(i)}\rangle$, we use
the Lagrange multiplier $\lambda$ to enforce the constraint
$\langle \Psi|\Psi\rangle=1$,
\begin{equation}
f(\Psi)\equiv \langle \Psi|g\rangle\langle g|\Psi\rangle -\lambda
\langle \Psi|\Psi\rangle.
\end{equation}
Maximizing $f$ with respect to the local product state
$|\psi^{(i)}\rangle$, we obtain the extremal condition
\begin{equation}
{\cal H}_{\rm eff}^{(i)}|\psi^{(i)}\rangle = \lambda N^{(i)}
|\psi^{(i)}\rangle,
\end{equation}
where ${\cal H}_{\rm eff}^{(i)}\equiv (\bigotimes_{j\ne
i}^{N}\langle\psi^{(j)}|)|g\rangle\langle g|(\bigotimes_{j\ne
i}^{N}|\psi^{(j)}\rangle)$ is proportional to a local projector
(labeled by $|\phi^{(i)}\rangle\langle\phi^{(i)}|$ at $i$-site),
and the normalization $N^{(i)}\equiv\bigotimes_{j\ne
i}^{N}\langle\psi^{(j)}|\psi^{(j)}\rangle$,
 is unity if all the local states are properly normalized. From the
viewpoint of the variational MPS, one fixes all local states
$|\psi^{(j)}\rangle$ but $|\psi^{(i)}\rangle$ and solves for the
corresponding optimal $|\psi^{(i)}\rangle$ and repeats the same
procedure for $i+1$, $i+2$, etc. until the $N$-th site and sweeps
the procedure back and forth until the eigenvalue $\lambda$
converges. The converged value $|\lambda|^2$ is the square of the
maximal overlap $\Lambda_{\max}^2$.

Experimentally, this procedure means that one chooses randomly the
local measurement basis and picks arbitrarily one direction (i.e.,
rank-one projector) for each site, say, $|\psi^{(j)}\rangle$ for
the $j$-th site and only varies the basis for one site, say,
$i$-th at a time with all others fixed until one reaches a basis
where the measurement outcome along one direction occurs with the
most probability. Then one moves to the next site, say, $(i+1)$-th
site, and finds the optimal direction and repeats this procedure
by sweeping back and forth until the probability for the most
likely outcome converges.

\begin{figure}
\includegraphics[width=3.5in]{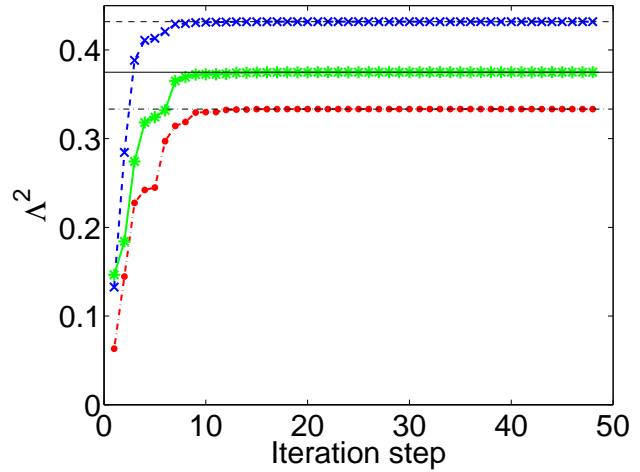}
\caption{(Color online). Numerical implementation of the iteration
procedure for the maximal overlap in the 4-spin XXZ chain. The
overlap square is shown as marked as ''*'', ''$\cdot$'' and
''$\times$'' for $\gamma=-0.9$, $1$ and $3$, respectively. An
initial 4-spin product state is randomly chosen. One round of
sweep starts from the first spin, reaches the forth spin, and
sweeps back to the second spin. One round thus contains six steps.
Eight iteration rounds are shown. The maximal overlap squares
converge rapidly to their exact values, which are indicated by the
thin lines, after just a few rounds. We have checked that the
converged product states are $|+-+-\rangle$ (up to a relative
phase), $|\phi\phi^\perp\phi\phi^\perp\rangle$ (with
$\langle\phi^\perp|\phi\rangle=0$), and $|0101\rangle$,
respectively.} \label{figit}
\end{figure}

Numerically, it appears that one has to solve the above eigenvalue
problem. But as the effective local Hamiltonian is a projector,
one immediately sees that the optimal rank-one projector is
exactly the one ($|\phi^{(i)}\rangle\langle\phi^{(i)}|$) given by
${\cal H}_{\rm eff}^{(i)}$. One thus replaces $|\psi^{(i)}\rangle$
by $|\phi^{(i)}\rangle$ and repeats the procedure at other sites
until the overlap converges. Experimentally, the apparatus setting
in line with the projector gives the local maximum output
probability. The search for the optimal direction at the $i$-th
site need not be a blind search, as a tomography (conditioned on
all other sites being measured in their respective
$|\psi^{(j)}\rangle$) will enable the determination of the
 optimal local direction $|\phi^{(i)}\rangle$. The whole procedure, either numerically or
experimentally, thus becomes an iterative procedure, given an
initial choice of $\{|\psi^{(j)}\rangle\}$. We have performed such
a numerical procedure and have found that this procedure converges
efficiently to the maximal overlap. The convergence for the ground
state of four-spin chain is very fast and the result is accurate;
see Fig.~\ref{figit}. In our experiment, we implement a simplified
version to directly measure the ground-state entanglement.

\section{Method for solving the 4-spin XXZ chain}


For $\gamma<-1$ the ground state  is doubly degenerate with ground
states being $|0000\rangle$ and $|1111\rangle$. To avoid the
complication due to the degeneracy, we can introduce an additional
small Zeeman term $H_{z} = B_{z}\sum_{i=1}^{N} Z_{i}$ with $0 <
B_z \ll 1$ into the Hamiltonian to lift the energy of
$|0000\rangle$ so that the ground state $|g\rangle$ becomes
$|1111\rangle$. The small Zeeman term does not affect the
universality class of phase transition of the ground state,
because it commutes with $H_{XXZ}$.

For $\gamma>-1$, the ground state is not a simple product state.
Due to the absence of an external field, the conservation of
z-component total angular momentum and the periodic translation
invariance give rise to only two relevant ``Bethe-ansatz'' basis
states for the ground states:
\begin{eqnarray}\label{bas1}
    |\phi_{1}\rangle &\equiv&\frac{1}{\sqrt{2}}(|0101\rangle+|1010\rangle)\\
\label{bas2}
    |\phi_{2}\rangle
    &\equiv&\frac{1}{2}(|1100\rangle+|0011\rangle+|1001\rangle+|0110\rangle).
\end{eqnarray}
By solving the effective Hamiltonian in the subspace spanned by
$\{|\phi_{1}\rangle,|\phi_{2}\rangle\}$
\begin{equation}\label{effH}
  H_{eff}=-4\left(%
\begin{array}{cc}
  \gamma & -\sqrt{2} \\
  -\sqrt{2} & 0 \\
\end{array}%
\right)
\end{equation}
we obtain that the ground state is
\begin{equation}\label{groundp}
    |g\rangle =|\phi_{1}\rangle \cos\alpha + |\phi_{2}\rangle\sin\alpha
\end{equation}
where $\alpha\in(-\pi/2,0)$ is given via
\begin{equation}\label{angle}
  \tan(2\alpha)=-2\sqrt{2}/\gamma,
\end{equation}
and that the ground energy is
\begin{equation}\label{Eg}
    E_g(\gamma) =-2\gamma-2\sqrt{\gamma^{2}+8}.
\end{equation}

\section{Method for experimental implementation}
\begin{figure}
\includegraphics[width=3.5in]{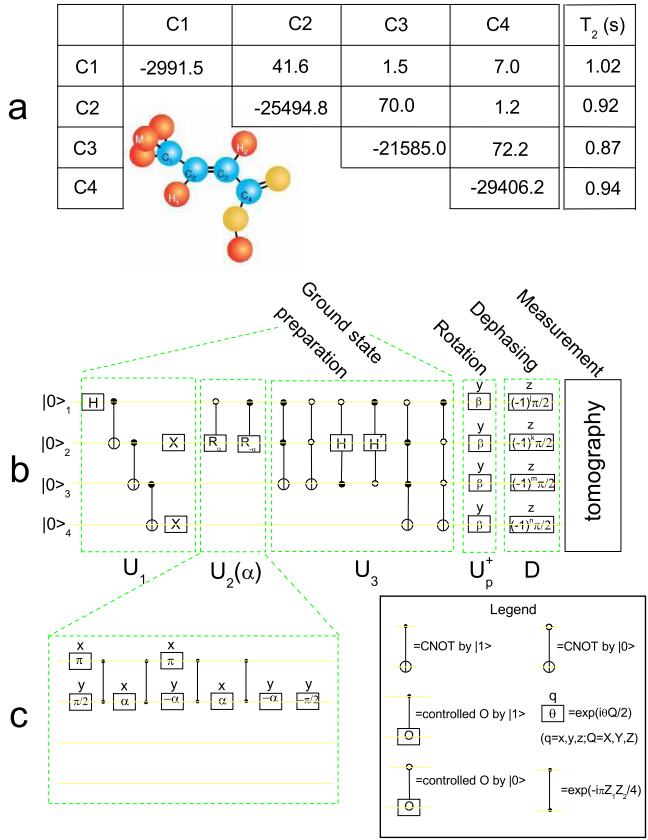} 
\caption{(Color online).  Experimental protocol. (a) The molecule
structure (inset) and the parameters of the four carbon-13 spins.
The diagonal terms show the chemical shifts 
 and the
non-diagonal terms show the strength of $J$- couplings 
in Hz.
 The transversal relaxation times $T_{2}$ measured by
a Hahn echo are listed in the rightmost column. (b) Quantum
circuit for creating the ground state and measuring its overlap
$\Lambda_{i}(\gamma)$ for $\gamma > -1$. 
 Here $\mathbf{X}$ denotes a
NOT gate. $R_{\pm\alpha}=\exp{(\pm i\alpha Y)}$, and $\mathbf{H}$
( $\mathbf{H^{*}}$) denotes a 
gate  transforming state $|0\rangle$ ($|1\rangle$) to $(|0\rangle$
+ $|1\rangle)/\sqrt{2}$, and $|1\rangle$ ($|0\rangle$) to
$(|0\rangle$ - $|1\rangle)/\sqrt{2}$, respectively. The other
gates are illustrated in the legend box. 
$U_{2}$ is decomposed
as the gate sequence shown in (c), 
for creating an arbitrary ground state for $\gamma > -1$ through
varying $\alpha$ [see Eq. (\ref{angle})]. $U_{p}$ denotes a
rotation for obtaining the overlap of the ground-state with
$U_{p}|b\rangle$, where $|b\rangle$ denotes a computational basis.
In dephasing operation, $j$, $k$, $m$, $n$ are chosen as $0$ and
$1$, respectively, 
to average out all non-diagonal terms in the density matrix to
zero. Tomograhpy of the diagonal terms requires four $\pi/2$
pulses selective for C1 to C4, respectively.} \label{figcir}
\end{figure}

   The experiment is performed in a Bruker DRX 700 MHz spectrometer.
   The structure of the molecule of crotonic acid and the parameters of the 
four spin qubits
   are shown in Fig. \ref{figcir} (a). The protons are decoupled in the whole experiment.
The initial pseudo-pure state $|0000\rangle$ is prepared  by
spatial averaging \cite{zhang09a,pures}, and chosen as the
reference state for normalizing the signals in the following
partial state tomography.

   The quantum circuit shown as Fig. \ref{figcir} (b) illustrates the
experiment protocol for $\gamma>-1$. The ground state is created
by $U_{3}U_{2}(\alpha)U_{1}$, indicated by the three dashed
blocks, respectively. We optimize  $U_{1}$ and $U_{3}$, which are
independent of $\alpha$, as two long (40 ms duration) GRAPE
pulses, respectively, where GRAPE stands for gradient ascent pulse
engineering.  The theoretical fidelity for $U_{1}$ and $U_{3}$ is
larger than $99\%$. To save time in searching GRAPE pulses,  we
further decompose $U_{2}(\alpha)$ into simple gates shown as the
sequence in Fig. \ref{figcir} (c), where each gate is implemented
by one GRAPE pulse. The duration of the pulse for the spin
coupling evolution is 20 ms, and the duration of other pulses is
0.5 ms. The theoretical fidelity for each pulse in Fig.
\ref{figcir} (c) is larger than $99.5\%$. An arbitrary ground
state for $\gamma>-1$ can be generated through varying $\alpha$ in
the single spin operation, which is much easier to find than
$U_{2}(\alpha)$ in the GRAPE algorithm. For the case of
$\gamma<-1$, we replace $U_{3}U_{2}(\alpha)U_{1}$ by four NOT
gates implemented by four $\pi$ pulses applied to the four qubits
respectively to create the ground state $|1111\rangle$ from
$|0000\rangle$.

To measure the overlap between $|g\rangle$ and an arbitrary
product state $|\Psi\rangle$, we re-write the overlap
$\Lambda\equiv\langle \Psi|g\rangle$
  in form of
\begin{equation}\label{overlapr}
    \Lambda=\langle b|U^{\dag}_{p}|g\rangle
\end{equation}
where $|b\rangle$ denotes a computational basis and
$U_{p}|b\rangle=|\Psi\rangle$ \cite{zhang09a}. Here we choose
$U_{p}$ as
\begin{equation}\label{SUpexp}
U_{p}(\beta)=\bigotimes_{j=1}^{4} e^{-i\beta Y_{j}/2}.
\end{equation}
 Since
$|\Psi_{1}\rangle$ and $|\Psi_{3}\rangle$ are already the
computational basis, we can simply choose $|b\rangle$ as
$|\Psi_{1}\rangle$ and $|\Psi_{3}\rangle$, respectively, and take
$U_{p}$ as the identity operation by setting $\beta=0$, for
obtaining $\Lambda_{1}$ and $\Lambda_{3}$ from Eq.
(\ref{overlapr}). $|\Psi_{2}\rangle$ is not a computational basis.
We can choose $|b\rangle=|0101\rangle$ and $\beta=\pi/2$ for
obtaining $\Lambda_{2}$, noting that
$|\Psi_{2}\rangle=U_{p}(\pi/2)|0101\rangle$.

In the density-matrix form, Eq. (\ref{overlapr}) is represented as
\begin{equation}\label{overlapdens}
    \Lambda^{2}= Tr(|b\rangle\langle b|\rho)
\end{equation}
where $\rho = U_{p}^{\dag}(|g\rangle \langle g|) U_{p}$. From Eq.
(\ref{overlapdens}), one finds that $\Lambda^{2}$ is encoded as
the diagonal element $|b\rangle\langle b|$ of $\rho$. We exploit
phase cycling to remove all the non-diagonal elements, and then
reconstruct all the diagonal terms of $\rho$ using partial state
tomographgy through four $\pi/2$ readout pulses selective for
C1-C4, respectively. $\Lambda^{2}$ is therefore extracted from the
diagonal terms.

\begin{thebibliography}{}
\bibitem{note1} Here $|+-+-...\rangle$ is
equivalent to $|-+-+...\rangle$ or more generally
$|\theta_+\theta_-\theta_+\theta_-...\rangle$ with
$\theta_\pm\equiv (|0\rangle \pm e^{i\theta}|1\rangle)/\sqrt{2}$
due to the symmetry in the ground states. Similarly,
$|1010...\rangle$ is equivalent to $|0101...\rangle$.
\end{thebibliography}
\end{document}